Field-Strength Descriptions for a System of Classical SU(2) Charges
with Spherical Symmetry and Confining Boundary Conditions.

Dennis Sivers

Portland Physics Institute
Portland, OR 97239

Spin Physics Center
University of Michigan
Ann Arbor, MI 48109

**Abstract**

The existence of a mechanism within the non-Abelian dynamics of QCD that confines quarks and gluons to the interior of hadrons has long been accepted empirically. To explore what this mechanism might look like, this paper examines field-strength descriptions for an extended system of SU(2) charges with spherical symmetry and imposes alternate confining boundary conditions to the time-independent Yang-Mills Maxwell equations. Three types of global solutions to the set of equations can be distinguished: types 0,1 and 2. Type-0 solutions evade the nonlinear dynamics associated with the radial magnetic field to describe a topologically trivial bound state. Type-1 and type-2 solutions both require a domain wall of topological charge to separate the interior volume containing the SU(2) charge densities from the exterior volume where the boundary conditions are imposed. Type-1 solutions describe an exterior volume with a radial magnetic field while type-2 solutions contain a sterile exterior volume where all field-strengths vanish. These solutions both describe a non-Abelian "spherical dual topological insulator". The dimensional reduction associated with the imposition of spherical symmetry on this system of SU(2) charges can also be applied to SU(3) charges so this simple exercise is directly relevant to understanding confinement in QCD.

Section 1.  Introduction

The intriguing concept that the confinement of color charges in QCD results from nonvanishing vacuum condensates leading to a non-Abelian manifestation of the dual-Meissner effect [1] for bulk superconductors can be motivated in many ways [2]. Early exploration of this important idea by 't Hooft [3] and Mandelstam [4] provided the initial creative impetus for significant theoretical work on this subject.  Importantly, the famous papers of Sieberg and Witten [5] provide an explicit example of the dual-Meissner effect in a quantum field theory.   The Seiberg-Witten papers demonstrate that the breaking of an N=2 supersymmetry in a supersymmetric Yang-Mills theory down to an N=1 supersymmetry in the presence of a small mass term for matter with an adjoint charge triggers a monopole or dyon condensation and leads directly to a U(1) confinement of trial charges.  Since the publication of these important results there has been an intense theoretical effort to extend the underlying ideas to help find possible condensation mechanisms for QCD charges that do not lead to the extra states implied by U(1) confinement [6] but display, instead, the N-ality and Casimir constraints found by lattice gauge theory simulations and strongly suggested by many other phenomenological arguments. [7]

The search for a particular mechanism leading to the vacuum condensation of mobile color-magnetic charges that could provide a dual equivalent to the Cooper-pair condensates for bulk BCS superconductors [8] has now been presented with a new type of challenge.  An important series of papers has provided compelling arguments [9,10] implying that relations involving phenomenological condensates of quarks and gluons, such as

$$\left\langle \frac{\alpha_S}{\pi} G^a_{\mu\nu} G^{\mu\nu a} \right\rangle = (0.012 GeV^4)(1\pm 0.3) \tag{1.1}$$

a parameter used in applications of QCD sum rules [11], along with the chiral condensate,

$$\left\langle \bar{q}_i q_i \right\rangle = -(0.24 \pm 0.01 GeV^3), \tag{1.2}$$

another parameter that provides a crucial component necessary for the understanding of dynamic chiral symmetry breaking [12], need not necessarily be interpreted as representing properties of the QCD vacuum but can instead be directly associated with the complex, nonperturbative dynamics found in the interior of hadrons.  The descriptive term "in-hadron condensates" has therefore been applied to this alternative approach to the theoretical scaffolding of dynamical symmetry breaking and to the techniques of summing over unseen states in QCD calculations. However, significant counter-arguments [13] have been raised in response to this recent interpretation of QCD condensates as localized hadronic properties rather than as translationally-invariant properties of the ground state of the underlying theory. Absence of vacuum condensates would also significantly impact the dual-Meissner approach to color confinement. The interplay of these ideas found in Refs. [9,10,13] seems certain to engender an ongoing constructive dialog leading to a thorough re-examination of many theoretical ideas involving models and mechanisms for both dynamical chiral symmetry breaking and for the phenomenological application of QCD sum rules.

This paper presents a modest example of this type of re-examination with the intent of extending the discussion of "in-hadron" condensates to clarify their challenge to the dual-Meissner explanation of color confinement. The example avoids formal consideration of the field-theoretical description of the vacuum and, instead, directly examines the spatial distribution of field-strength densities for a confined volume of non-Abelian charges. It is based on work the author first began in collaboration with John Ralston [14,15] and then pursued independently [16,17,18] which explored the various aspects of the dual-Meissner effect within the framework of a specific ansatz [18] for a the vector potential of an SU(2) gauge theory with spherical symmetry that was first proposed by Bars [19] and applied by Witten [20] to discuss multi-instanton configurations. As we will describe below, the imposition of spherical symmetry reduces the SU(2) Lagrangian density

$$\int d^4 x L^G(x) \Rightarrow 4\pi \int dr dt (r^2 L^G) \tag{1.3}$$

to a structure with

$$r^2 L^G = r^2 (F_{lm} F^{lm}) + 2 D^l \Phi D_l \Phi^* + \frac{1}{r^2} (|\Phi|^2 - 1)^2. \tag{1.4}$$

where the indices $l, m = 0, 1$ denote the variables $t, r$.

The form of the SU(2) gauge action with spherical symmetry, thus, replicates the Abelian Higgs/ Landau Ginzburg [21] action for a bulk superconductor in a 1+1 dimensional curved space-time. This identification implies that the Bars-Witten ansatz is well-suited to describing extended systems as well as permitting specific solutions representing non-Abelian exotica such as instantons, merons, monopoles and dyons [22-23]. In addition, this formalism can be augmented to include spherically-symmetric fermionic sources with an SU(2) Lagrange density

$$\int d^4 x L^F(x) \Rightarrow 4\pi \int dr dt (r^2 L^F) \tag{1.5}$$

with

$$r^2 L^F = r^2 \bar{\psi} (\slashed{D}_A - m) \psi \tag{1.6}$$

in a manner which fits appropriately into the two-dimensional curved space-time of (1.4). It is worth noting that the imposition of spherical symmetry and the dimensional reduction resulting from this ansatz is topologically safe since the two homotopy groups

$$\Pi_1(U_1) = \Pi_3(SU(2)) = Z \tag{1.7}$$

resulting from the imposition of boundary conditions in the Abelian 1+1 dimensional theory and the non-Abelian 3+1 dimensional theory both coincide with the group, Z, of integers under addition. As will be demonstrated below, there are compelling arguments that the field-theoretical system defined by (1.3)-(1.6) allows for confining solutions with a mass gap. The field-strength characterization of these classical solutions may therefore suggest a possible starting point for the construction of a color-confining quantum theory for QCD in 3+1 dimensions.

Examples of the results found in refs. [14-18] will be discussed below to introduce necessary tools for the characterization of non-Abelian field-strength densities. The initial studies have been successful in providing an elementary guide to understanding a possible background field approximation for modeling the medium effect of the QCD vacuum. However, when these techniques for the description of spherically symmetric systems began to be applied to other phenomenological issues of hadron structure [24-26] it became evident that a class of solutions to the field equations not adequately considered in [14-18] could play a role in explaining some puzzling aspects of nonperturbative dynamics in non-Abelian systems. These neglected solutions are best described in terms of their field-strength densities. As indicated in Fig. 1, these solutions describe an interior volume, $r \in [0, R_0 - \Delta)$, of color-charged matter density but vanishing topological charge surrounded by a shell volume, $r \in (R_0 - \Delta, R_0 + \Delta)$, of nonvanishing topological charge density, characterized by

$$E_i^a B_i^a(r) \neq 0, \tag{1.8}$$

and then completed with an open exterior region with infinite volume, $r \geq R_0 + \Delta$, where either there is a radially-directed magnetic field or where <u>all</u> of the field-strength densities vanish. In contrast to the expected sharp distinction of the two types of solutions with one showing a display of power by an exterior system with the bulk ensemble of mobile color-magnetic charges of a dual-Meissner vacuum, both types of solutions, instead, show the importance of a self-maintained layer non-Abelian dyonic charge. [27,28, 29] Both types of solutions are similar to condensed-matter systems displaying the properties of topological insulators have recently attracted intense theoretical and experimental interest [30-42]. It is therefore conceivable that the beginning of an explanation of color confinement in QCD can be found by carefully examining the idea that non-Abelian charges inside hadrons can be shielded by a shell of topologically charged shrink-wrap with mobile dyonic and topological charges. The topological charge density leading to (1.8) is, of course, notable for its association with the "strong-CP" problem in QCD [43-47]. The topological properties of non-Abelian dyonic charges have led to the prediction of an axion particle as the Goldstone boson of a broken Peccei-Quinn symmetry of the strong interactions [48]. It is the consequence of the CP-odd nature of this topologically-charged layer that provides the potential capability for it to isolate color charges into color-singlet states.

The remainder of this paper is organized as follows. Section 2 summarizes the Bars-Witten [19,20] parameterization of an SU(2) gauge connection using the form of the basis tensors with both spatial and gauge indices as applied by Ralston and Sivers [14-16]. The advantage of the field-strength tensor in describing extended systems is also discussed here. Section 3 introduces the Weyl-Dirac equations for fermions in a fundamental representation of SU(2) and briefly discusses properties of the combined system.[17-18] In Section 4 we present simple static solutions to the combined system using classical sources with confining boundary conditions using the tools implicit in the form of the Lagrangian presented in (1.4). As indicated above, we consider three types of confining boundary conditions: types 0,1 and 2. Type-0 solutions are topologically trivial while the other two types require the existence of an emergent structure consisting of a domain wall of spherically-symmetric topological charge separating the interior and exterior volumes with type 1 representing and exterior volume with a radial magnetic field and type 2 representing a sterile exterior volume where all sources and field strengths vanish. Section 5 briefly describes some of the properties of the transition region associated with the gradients required by the three types of boundary conditions while Section 6 gives an outline for using these solutions to construct a quantum theory.

Section 2. A field-strength description of the Bars-Witten ansatz for SU(2).

One of the most useful *Ansatz* for an SU(2) gauge theory involves the assumption that all observables display a spherical symmetry. This assumption was originally incorporated into a parameterization of the gauge potential for a 4-dimensional Euclidean formulation of SU(2) pioneered by Bars [19] and Witten [20]. For this paper, we will be working in Minkowski space with metric diag (-1,1,1,1) and will parameterize the SU(2)-valued vector potential in the form [15,16],

$$gA_0^a(r,t) = A_0(r,t)\hat{r}_a$$
$$gA_i^a(r,t) = A_1(r,t)\rho_{ia} + \frac{a(r,t)}{r}\sin\omega(r,t)\delta_{ia}^T + \frac{a(r,t)\cos\omega(r,t)-1}{r}\varepsilon_{ia}^T, \quad (2.1)$$

where we have absorbed the gauge coupling, g, into the vector potential. The underlying tensor structure of this equation can be understood in terms of the diagram shown in Fig. 2. The tensors $\rho_{ia}$, $\delta_{ia}^T$ and $\varepsilon_{ia}^T$ combine spatial indices $i,j,k=1,2,3$ to adjoint SU(2) group indices $a,b,c=1,2,3$ and can be written in the form

$$\rho_{ia} = \hat{r}_i \hat{r}_a$$
$$\delta^T_{ia} = \delta_{ia} - \rho_{ia} = (\hat{\theta}_i \hat{\theta}_a + \hat{\phi}_i \hat{\phi}_a) \quad (2.2)$$
$$\varepsilon^T_{ia} = \varepsilon_{ial} \hat{r}_l = (\hat{\phi}_i \hat{\theta}_a - \hat{\theta}_i \hat{\phi}_a)$$

In this parameterization the traces for the transverse tensors, $\delta^T_{ia}\delta^T_{ia} = 2$ and $\varepsilon^T_{ia}\varepsilon^T_{ia} = 2$, differ from the trace of the longitudinal tensor, $\rho_{ia}\rho_{ia} = 1$. The transverse tensors (2.2) simplify the form of the non-Abelian field equations in view of the algebraic relations,

$$\frac{1}{r}\delta^T_{ia} = \partial_i \hat{r}_a$$
$$\frac{1}{r}\varepsilon^T_{ia} = -i[\hat{r}, \partial_i \hat{r}]_a \quad (2.3)$$

where $[x_a, x_b] = i\varepsilon_{abc} x_c$ is the SU(2) commutator. Because the non-Abelian field equations involve the vector potential through the gauge-covariant derivative,

$$D^{ab}_\mu = \partial_\mu \delta^{ab} + g\varepsilon^{abc} A^c_\mu, \quad (2.4)$$

we can combine (2.1) and (2.3) to define the gauge-dependent transverse tensors $e^S_{ia}(\omega)$ and $e^A_{ia}(\omega)$ given by

$$(D^{ab}_i \hat{r}_b) = \frac{a(r,t)}{r} e^S_{ia}(\omega(r,t))$$
$$-i[\hat{r}, D_i \hat{r}]_a = \frac{a(r,t)}{r} e^A_{ia}(\omega(r,t)) \quad (2.5)$$

where

$$e^S_{ia}(\omega) = \delta^T_{ia} \cos(\omega) - \varepsilon^T_{ia} \sin(\omega)$$
$$e^A_{ia}(\omega) = \delta^T_{ia} \sin(\omega) + \varepsilon^T_{ia} \cos(\omega). \quad (2.6)$$

The superscripts A and S designate antisymmetric or symmetric behavior under the interchange of spatial and SU(2) color indices,

$$e^S_{ai}(\omega) = e^S_{ia}(-\omega)$$
$$e^A_{ai}(\omega) = -e^A_{ia}(-\omega). \quad (2.7)$$

The two transverse tensors can also be related by a shift in the gauge angle $\omega(r,t)$, using the expressions

$$e^S_{ia}(\omega + \frac{\pi}{2}) = -e^A_{ia}(\omega)$$
$$e^A_{ia}(\omega + \frac{\pi}{2}) = e^S_{ia}(\omega), \quad (2.8)$$

and we have the multiplicative relations

$$e^S_{im}(\omega_1)e^S_{ma}(\omega_2) = e^S_{ia}(\omega_1+\omega_2)$$
$$e^S_{im}(\omega_1)e^A_{ma}(\omega_2) = e^A_{ia}(\omega_1+\omega_2) \quad (2.9)$$
$$e^A_{im}(\omega_1)e^A_{ma}(\omega_2) = -e^S_{ia}(\omega_1+\omega_2).$$

These relations demonstrate that the transverse tensors provide a representation of the group U(1) traditionally associated with the complex phase $\exp(i\omega)$ combined with the use of the interchange of spatial and color indices to correspond with complex conjugation.

At this stage it is convenient to write the field-strength tensor

$$G^a_{\mu\nu} = \partial_\mu A^a_\nu - \partial_\nu A^a_\mu + g\varepsilon^{abc} A^b_\mu A^c_\nu \quad (2.10)$$

in terms of the electric and magnetic fields $E^a_i = G^a_{0i}$ and $B^a_i = \frac{1}{2}\varepsilon^{ijk} G^a_{jk}$,

$$gE^a_i = E_L(r,t)\rho_{ia} + E_S(r,t)e^S_{ia}(\omega(r,t)) + E_A(r,t)e^A_{ia}(\omega(r,t)), \quad (2.11)$$

with electric field-strength components written as

$$E_L(r,t) = -\partial_t A_1(r,t) + \partial_r A_0(r,t)$$
$$E_S(r,t) = \frac{a(r,t)}{r}[-\partial_t \omega(r,t) + A_0(r,t)] \quad (2.12)$$
$$E_A(r,t) = \frac{-\partial_t a(r,t)}{r}$$

and

$$gB^a_i = B_L(r,t)\rho_{ia} + B_S(r,t)e^S_{ia}(\omega(r,t)) + B_A(r,t)e^A_{ia}(\omega(r,t)) \quad (2.13)$$

with magnetic field-strength components given by

$$B_L(r,t) = \frac{a^2(r,t)-1}{r^2}$$
$$B_S(r,t) = -\frac{\partial_r a(r,t)}{r} \quad (2.14)$$
$$B_A(r,t) = \frac{a(r,t)}{r}[A_1(r,t) - \partial_r \omega(r,t)]$$

To study the field-strength description of an extended system, it is important to note that the gauge covariant derivatives in (2.5) are proportional to $a(r,t)$ and that any nonlinearity found in the definition of the field-strength density is necessarily associated with the specification of the $\hat{r}$-directed magnetic field, $B_L(r,t)$. We can also define the 2-dimensional topological current involving the gauge potential (2.1),

$$K_0(r,t) = (a^2(r,t)-1)A_1(r,t) - a^2(r,t)\partial_r\omega(r,t)$$
$$K_1(r,t) = -(a^2(r,t)-1)A_0(r,t) + a^2(r,t)\partial_t\omega(r,t), \qquad (2.15)$$

such that,

$$\partial^l K_l(r,t) = \frac{1}{2}g^2 r^2 G^{*a}_{\mu\nu} G^{a\mu\nu} = g^2 r^2 E_i^a B_i^a. \qquad (2.16)$$

In the non-Abelian generalization of the Maxwell equations,

$$(D^\mu G_{\mu\nu})^a = J_\nu^a(r,t) \qquad (2.17)$$

with the parameterization of the classical currents given by,

$$J_0^a(r,t) = \frac{1}{r^2} J_0(r,t)\hat{r}_a$$
$$J_i^a(r,t) = \frac{1}{r^2} J_1(r,t)\rho_{ia} + j_S(r,t)\varepsilon_{ia}^S(\omega) + j_A(r,t)\varepsilon_{ia}^A(\omega), \qquad (2.18)$$

,

the SU(2) Yang-Mills Maxwell equations become

$$-\partial_r(r^2 E_L(r,t)) + 2ar E_S(r,t) = qJ_0(r,t)$$
$$-\partial_t(r^2 E_L(r,t)) + 2ar B_A(r,t) = qJ_1(r,t)$$
$$-\partial_t(ar E_S(r,t)) + \partial_r(ar B_A(r,t)) = qar j_S(r,t) \qquad (2.19)$$
$$a(-\partial_t^2 + \partial_r^2)a(r,t) - r^2(E_S^2 - B_A^2)(r,t) - \frac{a^2(a^2-1)(r,t)}{r^2} = qar j_A(r,t).$$

In these classical equations, the charge "q" on the RHS absorbs the normalization of the currents. The non-Abelian generalization of the Bianchi constraints gives the additional equations,

$$\partial_r(ar E_A(r,t)) - \partial_t(ar B_S(r,t)) = 0$$
$$-E_L + \partial_r(ar E_S(r,t)) + \partial_t(ar B_A(r,t)) = \partial^l K_l = \frac{1}{2}g^2 r^2 G^{*a}_{\mu\nu} G^{a\mu\nu}. \qquad (2.20)$$

There are many examples of solutions to the equations (2.19) and (2.20) that are extensively discussed in references 14-16. We have chosen to express the equation for $j_A(r,t)$ as an equation directly written in terms of $a(r,t) = \pm(r^2 B_L(r,t)+1)^{\frac{1}{2}}$ instead of $B_L(r,t)$ itself. Note that every expression in (2.19) and (2.20) includes a factor of $a(r,t)$ and that all the nonlinearity in these equations is encapsulated in the expression for $j_A(r,t)$. We can identify

$$|a(r,t)| = |\Phi| \qquad (2.21)$$

in the Lagrange density defined in (1.3) and (1.4) with the transverse field densities acting like charged scalars in 1+1 dimensions. In Section 4, we will be looking at static kink solutions to (2.19) with boundary conditions chosen to enforce confinement of SU(2) charge. Before doing this, however, it is convenient to introduce SU(2) fermions into the formalism. This will provide some additional meaning to the topological current in (2.15) and (2.16)

## Section 3. SU(2) Fermions and Currents in the non-Abelian Maxwell's equations.

The correspondence between the 3+1 dimensional SU(2) gauge theory with spherical symmetry and the 1+1 dimensional Abelian Higgs model in curved space can be strengthened by the introduction of SU(2) fermions in the fundamental representation. Based on the formalism presented in Refs. 17 and 18 we can parameterize $\hat{r}$-directed Weyl spinors, $\xi_I^\pm$, in 3-space and corresponding spinors, $\eta_B^\pm$, such that

$$\sigma_I^{rJ} \xi_J^\pm = \pm \xi_I^\pm \quad I, J = 1, 2 \tag{3.1}$$

and

$$\tau_B^{rC} \eta_C^\pm = \pm \eta_B^\pm \quad B, C = 1, 2 \tag{3.2}$$

where $\sigma^r = \sigma \cdot \hat{r}$ and $\tau^r = \tau \cdot \hat{r}$ are, respectively the $2 \times 2$ Pauli matrices in 3-space and group space. Bispinors of the form

$$X_{IB}^{\pm\pm}(\theta, \phi) = \xi_I^\pm \eta_B^\pm \tag{3.3}$$

with one spatial index and SU(2) index define appropriate objects for describing the set of spherically symmetric SU(2) fermion fields in the fundamental representation. Thus, they play a role in defining a basis for fermions analogous to the basis $\rho_{ia}, \delta_{ia}^T$ and $\varepsilon_{ia}^T$ in defining a basis for gauge-covariant 3-vectors in the adjoint representation of SU(2). The Dirac equation can be decomposed into two Weyl equations

$$\sigma_{IJ}^\mu (\partial_\mu - ig A_\mu^a (\frac{\tau^a}{2})_B^C) \psi_{RI}^{\;\;J} = m \psi_{LIB}$$

$$\sigma^{\mu IJ} (\partial_\mu - ig A_\mu^a (\frac{\tau^a}{2})_B^C) \psi_{LJC} = -m \psi_R^{\;\;I}{}_C \tag{3.4}$$

where m is the quark mass. The Weyl spinors can be further decomposed

$$\psi_R{}^J{}_C = \frac{1}{r}[R^+(r,t)X^{++}(\theta,\phi)^J_C + R^-(r,t)X^{--}(\theta,\phi)^J_C]$$

$$\psi_{LIB} = \frac{1}{r}[L^+(r,t)X^{-+}(\theta,\phi)_{IB} + L^-(r,t)X^{+-}(\theta,\phi)_{IB}].$$

(3.5)

To express the derivatives in (3.4) we can display them in spherical coordinates

$$\sigma^\mu \partial_\mu = -\sigma^0 \partial_t + \sigma^{\hat{r}} \partial_r + \sigma^{\hat{\theta}} \frac{\partial_\theta}{r} + \sigma^{\hat{\phi}} \frac{\partial_\phi}{r \sin\theta}. \tag{3.6}$$

Combining (3.5) with the parameterization of the gauge potential given in (2.1) leads to the set of equations [18]

$$\frac{1}{r}[-D_t - D_r]R^+ + \frac{\Phi}{r^2}R^- = \frac{m}{r}L^+$$

$$\frac{1}{r}[D_t - D_r]R^- + \frac{\Phi^*}{r^2}R^+ = \frac{m}{r}L^-$$

$$\frac{1}{r}[-D_t + D_r]L^+ + \frac{\Phi}{r^2}L^- = \frac{-m}{r}R^+$$

$$\frac{1}{r}[D_t + D_r]L^- + \frac{\Phi^*}{r^2}L^+ = \frac{-m}{r}R^-.$$

(3.7)

In writing (3.6) we have used the two-dimensional Abelian covariant derivative

$$D_l = \partial_l - ieA_l \tag{3.8}$$

and have made the corresponding identification of an imbedded Abelian charge for the fermion such that

$$e\eta_B^\pm = \pm \frac{1}{2}\eta_B^\pm. \tag{3.9}$$

The transverse derivatives can be simplified using the raising and lowering operators

$$\sigma^\pm = \frac{1}{2}(\sigma^{\hat{\theta}} \pm i\sigma^{\hat{\phi}}). \tag{3.10}$$

These operators have the action on the spatial spinors $\xi^\pm$,

$$\sigma^+ \xi^+ = 0 \quad \sigma^- \xi^+ = \xi^-$$
$$\sigma^+ \xi^- = \xi^+ \quad \sigma^- \xi^- = 0.$$

(3.11)

so that the transverse covariant derivative acts on the charge space,

$$\slashed{D}_T \xi_I^+ \eta_A^- = \slashed{D}_T \xi_I^- \eta_A^+ = \frac{1}{r}(\xi_I^+ \eta_A^- - \xi_I^- \eta_A^+). \tag{3.12}$$

A more complete set of the spinor conventions leading to (3.7) can be found the appendices of ref. [18].

We can further enhance the connection the Bars-Witten ansatz for the SU(2) gauge fields to the Abelian Higgs model in curved space by introducing a convenient basis for the 1+1 dimensional $\gamma$ matrices,

$$\gamma_0^{(2)} = r\begin{pmatrix} 0 & 1 \\ 1 & 0 \end{pmatrix}$$
$$\gamma_1^{(2)} = r\begin{pmatrix} 0 & -1 \\ 1 & 0 \end{pmatrix},$$
(3.13)

such that,

$$\{\gamma_l^{(2)}, \gamma_m^{(2)}\} = 2r^2 g_{lm}^{(2)} = 2g_{lm}^{(2\_curved)}.$$
(3.14)

With this representation of the $\gamma$ matrices we can define a 2-dimensional chirality operator

$$\gamma_5^{(2)} = \frac{1}{r^2}\gamma_0^{(2)}\gamma_1^{(2)} = \begin{pmatrix} 1 & 0 \\ 0 & -1 \end{pmatrix},$$
(3.15)

and introduce the 2-spinors

$$R = \begin{pmatrix} R^- \\ -R^+ \end{pmatrix} \quad L = \begin{pmatrix} L^+ \\ L^- \end{pmatrix}.$$
(3.16)

We can then define an effective pseudoscalar field formed from the transverse gauge fields by

$$\Phi^{(2)} = \frac{1}{2}\Phi(1+\gamma_5^{(2)}) + \frac{1}{2}\Phi^*(1-\gamma_5^{(2)}) = a(r,t)e^{i\omega(r,t)\gamma_5^{(2)}}.$$
(3.17)

Using (3.8) through (3.13) we can re-write (3.6) in the form

$$[\gamma_l^{(2)} D^l + \gamma_5^{(2)}\Phi^{(2)}(r,t)]R(r,t) = mL(r,t)$$
$$[\gamma_l^{(2)} D^l + \gamma_5^{(2)}\Phi^{(2)*}(r,t)]L(r,t) = -mR(r,t).$$
(3.18)

We will omit further discussion of the Weyl-Dirac equation for SU(2) fermions at this stage in order to consider the implications of artificially enforcing confinement of SU(2) color using specific boundary conditions on the SU(2) gauge fields with classical sources. The ability to deal with fermions within this ansatz remains an important feature both for the formalism and for the understanding of confinement.

Section 4. Static solutions to the non-Abelian Maxwell's equations with confining boundary conditions.

The discussion in Sections 2 and 3 has introduced the basic tools for understanding the classical field-strength densities within the Bars-Witten ansatz for spherically symmetric SU(2). In this section, we would like to use these tools to describe the properties of a static system of confined charges. Static solutions to the classical field equations involve the time independent field strengths,

$$E_L(r) = A_0'(r) \qquad E_S(r) = \frac{a(r)A_0(r)}{r} \qquad E_A(r) = 0$$
$$B_L(r) = \frac{a^2(r)-1}{r^2} \qquad B_S(r) = \frac{a'(r)}{r} \qquad B_A(r) = \frac{a(r)}{r}[A_1(r) - \omega'(r)]. \tag{4.1}$$

Note that in this section r-derivatives are written in the form $\frac{df(r)}{dr} = f'(r)$ and will take the gauge coupling, g, and the current charge, q, to both be unity. The static form of the Yang-Mills Maxwell equations for the Bars-Witten ansatz with classical currents can then be written,

$$-[r^2 E_L(r)]' + 2ra(r)E_S(r) = J_0(r)$$
$$2ra(r)B_A(r) = J_1(r)$$
$$[ra(r)B_A(r)]' = ra(r)j_S(r) \tag{4.2}$$
$$a(r)a''(r) - r^2[E_S^2(r) - B_A^2(r)] - \frac{a^2(r)[a^2(r)-1]}{r^2} = ra(r)j_A(r).$$

The static version of the topological current, (2.15) takes the form

$$K_0(r) = [a^2(r)-1]A_1(r) - a^2(r)\omega'(r)$$
$$K_1(r) = [1-a^2(r)]A_0(r) \tag{4.3}$$

with

$$K_1'(r) = -E_L(r) + [ra(r)E_S(r)]' \tag{4.4}$$

We would now like to consider a very simple set of time-independent solutions to (2.18) and (2.19) for an "interior volume" of 3-space defined by $r \leq R_0 - \Delta$. This solution will be described by the parameters,

$$A_0(r) = C_E r \qquad A_1(r) = C_M r$$
$$a(r) = 1 \qquad \omega(r) = 0. \tag{4.5}$$

In this parameterization $C_E$ and $C_M$ are constants when the equations are restricted to the interior volume.

These parameters lead to the static field strengths for the interior volume of 3-space defined by $r \leq R_0 - \Delta$,

$$E_L(r) = C_E \quad E_S(r) = C_E \quad E_A(r) = 0$$
$$B_L(r) = 0 \quad B_S(r) = 0 \quad B_A(r) = C_M. \quad (4.6)$$

The classical SU(2) color currents are specified by the components,

$$J_0(r) = 0 \quad J_1(r) = 2rC_M$$
$$rj_S(r) = C_M \quad rj_A(r) = r^2(C_M^2 - C_E^2) \quad (4.7)$$

The nontrivial Bianchi constraint takes the form,

$$-C_E + \frac{\partial}{\partial r}(rC_E) = \partial^l K_l = 0, \quad (4.8)$$

and the covariant conservation of the SU(2) color current

$$\frac{\partial}{\partial r} J_1(r) = 2rj_S(r)$$
$$2C_M = 2C_M$$

is also satisfied. The invariants,

$$E_i^a E_i^a - B_i^a B_i^a = E_L^2 - B_L^2 + 2(E_S^2 + E_A^2 - B_S^2 - B_A^2) = 3C_E^2 - 2C_M^2.$$
$$E_i^a B_i^a = E_L B_L + 2(E_S B_S + E_A B_A) = 0, \quad (4.9)$$

therefore define the interior volume to have a nonvanishing energy density but to display "vacuum" quantum numbers, $J^{PC} = 0^{++}$, for the SU(2) gauge fields so that the overall quantum numbers of the interior state can be defined in terms of the fermion content of the charged system. Notice that the classical currents do not depend on the sign of $C_E$. This is because of the reflection asymmetry $r \leftrightarrow -r$ embedded in the hypothesis of spherical symmetry. That symmetry can be broken by choosing the $\hat{r}$-directed static Poynting vector, $P_i = (E_S B_A - E_A B_S)\hat{r}_i$, to be $P_i = C_E C_M \hat{r}_i$. The two constants, $C_E$ and $C_M$, specify an extremely simple extended system of SU(2) color charge, but one with enough content to illustrate some of the impact of confining boundary conditions. For this purpose, we will consider three separate classes of confining boundary conditions defined by specifiying the field densities and color currents on the "exterior volume" defined by $r > R_0 + \Delta$. These three types of solutions are distinguished by the value of $a(r)$ in the exterior volume,

$$\text{type}-0 \quad a(r)=1$$
$$\text{type}-1 \quad a(r)=0$$
$$\text{type}-2 \quad a(r)=-1.$$

For each type of boundary condition we will interpolate the fields using the classical field equations through the shell volume defined by $r = R_0 + z\Delta$ with $z \in [-1,1]$.

<u>Type 0 boundary conditions.</u> We consider the vector potential (modulo gauge transformations) in the exterior volume region to be given by

$$a(r) = 1,$$
$$A_0(r) = A_1(r) = 0, \tag{4.10}$$

along with

$$\omega(r) = 0,$$
$$K_0(r) = K_1(r) = 0. \tag{4.11}$$

These boundary conditions guarantee that all of the field strengths are zero in the exterior volume. Consistency requires that he currents will also be set to zero,

$$J_0(r) = 0 \quad J_1(r) = 0 \quad j_S(r) = 0 \quad j_A(r) = 0 \tag{4.12}$$

We will impose these boundary conditions in a continuous form by specifying that the constants $C_E, C_M$ that define the charges in the interior region be multiplied by a function that interpolates between the values 1 and 0 through the transition region. The interpolation is, of course, not unique but in this simple example we will choose the set of functions,

$$\beta_\kappa(r) = \frac{1}{2}[1 - \tanh(\kappa z)], \tag{4.13}$$

with $z = (r - R_0)/\Delta$ and $\kappa$ chosen large enough so that

$$|\tanh(\kappa) - 1| \leq \varepsilon \tag{4.14}$$

for some small value of $\varepsilon$. Since we will be applying the static Yang-Mills Maxwell equations to the study of the transition region we will also require the derivatives

$$\beta_\kappa'(r) = -\frac{1}{2} \frac{\kappa}{\Delta \cosh^2(\kappa z)}$$
$$\beta_\kappa''(r) = \frac{\kappa^2 \tanh(\kappa z)}{\Delta^2 \cosh^2(\kappa z)} \tag{4.15}$$

It is convenient to also impose the additional constraints

$$\frac{\kappa}{R_0} \geq C_E \quad \frac{R_0}{\Delta} \geq C_E \quad \frac{\kappa}{\Delta} \geq C_E^2. \tag{4.16}$$

These constraints ensure that the ratio, $\kappa/\Delta$, that occurs in the derivatives of the interpolating functions be sufficiently large to allow for smoothness in the classical equations.

Type 1 boundary conditions. We also consider, for purposes of illustration, a "magnetic vacuum" with $a(r) = 0$ in the exterior volume and the field strengths given by

$$E_L(r) = 0, \quad E_S(r) = 0, \quad E_A(r) = 0, \tag{4.17}$$

and

$$B_L(r) = \frac{-1}{r^2}, \quad B_S(r) = 0, \quad B_A(r) = 0. \tag{4.18}$$

With $a(r) = 0$, the function $\omega(r)$ does not appear in the expression for the vector potential and we can specify that the color currents vanish

$$\begin{aligned} qJ_0(r) &= 0, \quad qJ_1(r) = 0, \\ qrj_S(r) &= 0, \quad qrj_A(r) = 0. \end{aligned} \tag{4.19}$$

With $a(r) = 0$, we have the topological currents given by

$$K_0(r) = -A_1(r), \quad K_1(r) = A_0(r), \tag{4.20}$$

and we can specify that both these Abelian components of the vector potential vanish in the exterior volume $r \geq R_0 + \Delta$,

$$A_0(r) = 0, \quad A_1(r) = 0, \tag{4.21}$$

leaving the surviving component of the SU(2) vector potential to be given by

$$gA_i^a(r) = \frac{-1}{r}\varepsilon_{ia}^T = \frac{-1}{r}\varepsilon_{ial}\hat{r}_l. \tag{4.22}$$

As with the case of type-0 boundary conditions, this expression in the exterior volume can, of course be further modified by gauge transformations but we will not consider the issues involving gauge transformations here. The exterior volume in this case has the behavior of a magnetic pole. Because the transverse components of the vector potential that carry color SU(2) charge vanish with $a(r) = 0$, color electric charge is automatically set to zero in the exterior region with these boundary conditions. In this set of solutions, the exterior region does not represent a classical vacuum state but, instead, can be characterized as the precursor for a simple model of a magnetic condensate in the exterior volume. To interpolate between the interior region and the exterior volume for these solutions, we will follow the same continuations for the type-0 solutions except that for $z \in (-1,1)$ we will use the continuation

$$a(r) = a_0 \beta_\kappa(r) \tag{4.23}$$

with $a_0 = 1$.

Type 2 Boundary Conditions.  The final set of boundary conditions that, like type-0 conditions, also to a "static sterile vacuum" in the exterior vacuum involve the requirement that $a(r) = -1$ for the region $r > R_0 + \Delta$. In addition to this condition, the electric and magnetic field strengths are specified to be:

$$E_L(r) = 0, \quad E_S(r) = 0, \quad E_A(r) = 0, \tag{4.24}$$

and

$$B_L(r) = 0, \quad B_S(r) = 0, \quad B_A(r) = 0. \tag{4.25}$$

The vanishing of the color currents in the exterior region leads to

$$\begin{aligned} qJ_0(r) = 0, \quad qJ_1(r) = 0, \\ qrj_S(r) = 0, \quad qrj_A(r) = 0, \end{aligned} \tag{4.26}$$

which duplicates the condition (4.19) for the type 1 boundary conditions. With $a(r) = -1$, the vanishing of $E_S(r)$ requires the gauge invariant constraint

$$-A_0(r) + K_1(r) = 0, \tag{4.27}$$

which can be satisfied by taking each of the terms to vanish in the static limit. The vanishing of $B_A(r)$ also gives a gauge-invariant condition

$$A_1(r) + K_0(r) = A_1(r) - \frac{\partial \omega(r)}{\partial r} = 0 \tag{4.28}$$

Both terms in this equation can be modified by an r-dependent gauge transformation such that

$$A_1' = A_1 + \frac{\partial \theta(r)}{\partial r} \quad K_0' = K_0 - \frac{\partial \theta(r)}{\partial r} \quad \omega' = \omega + \theta(r) \tag{4.29}$$

As can be seen directly from eq. (2.1), with $a(r) = -1$, the vector potential can only vanish for

$$\cos(\omega(r)) = -1, \quad \omega(r) = (2N+1)\pi \tag{4.30}$$

To match these conditions with the field strengths at $r = R_0 - \Delta$ we will follow the same rule for interpolation for the constants $C_E$ and $C_M$, as in the previous sections,

$$C_E \to C_E \beta_\kappa(r) \quad C_M \to C_M \beta_\kappa(r) \tag{4.31}$$

for type-0 and type-1 boundary conditions. However, for interpolating explicit factors of $a(R_0 + z\Delta)$ we will use

$$a(r) = a_0 \alpha_\kappa(r)$$
$$\alpha_\kappa(r) = -\tanh(\kappa z) \tag{4.32}$$

with $a_0 = 1$ and $\kappa$ again chosen large enough to give $\tanh(\pm\kappa) = \pm 1$. This changes the interpolation of the field strengths in one, very notable way. The covariant derivatives that define the basis for the transverse charge space defined in (2.5) change sign at $z = 0$ or $r = R_0$ within this type of boundary conditions. In addition, we will use the interpolation for $\omega(r)$ given by

$$\omega(r) = \pi \beta_\kappa(-r) \tag{4.33}$$

that interpolates between 0 and $\pi$ through the transition region.

The parameterization for the interpolations considered here is, in no way, unique. In addition the simple system of static SU(2) charges used to demonstrate the impact of confining boundary conditions is very artificial. What is notable here is that the three different sets of boundary conditions each lead to different, complicated charge structure in the transition region corresponding to $R_0 \pm \Delta$. These complicated charge structures can be illustrated by considering the ratio of the invariants $E_i^a E_i^a - B_i^a B_i^a$ and $E_i^a B_i^a$ for the three different types of solutions with each interpolation specified by the same parameters, $\Delta, \kappa, C_E, C_M$.

The most notable feature is that type-0 solutions involve no topological charge while for type-1 and type-2 solutions, the interpolations require that a layer of topological charge be generated by the non-Abelian Maxwell's equations. This layer shares many of the properties found in the topological insulators in condensed matter systems. [27-31]. These consequences will be described in Section 5.

Section 5. Field-Strength Comparisons in the Transition Region.

After specifying the three distinct types of confining boundary conditions and the requirements they impose for interpolating field-strength densities between the interior volume where color charges are specified, it is interesting to examine the different field strength densities in the transition region given by

$$r \in (R_0 - \Delta, R_0 + \Delta), \tag{5.1}$$

where there are, necessarily, color gradients. For the type-0 solutions with the interpolating functions given in Section 4, the SU(2) color field-strength densities are given by the expressions,

$$\begin{aligned} E_L(r) &= C_E[\beta_\kappa(r) + r\beta_\kappa'(r)] & E_S(r) &= C_E \beta_\kappa(r) & E_A(r) &= 0 \\ B_L(r) &= 0 & B_S(r) &= 0 & B_A(r) &= C_M \beta_\kappa(r). \end{aligned} \tag{5.2}$$

The transition region for type-0 solutions can therefore be described by the invariant field-strength densities,

$$(E_i^a E_i^a - B_i^a B_i^a)^{(0)}(r) = C_E^2([\beta_\kappa(r) + r\beta_\kappa'(r)]^2 + 2\beta_\kappa^2(r)) - 2C_M^2 \beta_\kappa^2(r)$$
$$(E_i^a B_i^a)^{(0)}(r) = 0. \tag{5.3}$$

A sketch of the scalar invariant density defined by (5.3) for the parameters $C_E = 3$, $C_M = 2$, $\kappa/R_0 = 3$ and $R_0/\Delta = 5$ is shown in Fig. (2). The striking spike in the color density found in the transition region is very sensitive to the ratio $\kappa/\Delta$ that appears in the expression for $\beta_\kappa'(r)$. There is no topological structure in type-0 solutions since the pseudoscalar color density vanishes. This suggests that the stability of these solutions is questionable.

The situation for the type-1 solutions is significantly different. The field-strengths in the transition region with the type-1 interpolating functions defined in Section 4 are:

$$E_L(r) = C_E[\beta_\kappa(r) + r\beta_\kappa'(r)] \quad E_S(r) = C_E \beta_\kappa^2(r) \quad E_A(r) = 0$$
$$B_L(r) = \frac{\beta_\kappa^2(r) - 1}{r^2} \quad\quad B_S(r) = \frac{-\beta'(r)}{r} \quad B_A(r) = C_M \beta_\kappa^2(r). \tag{5.4}$$

The transition region for the type-1 solutions can therefore be given in terms of the invariant densities:

$$(E_i^a E_i^a - B_i^a B_i^a)^{(1)}(r) = C_E^2([\beta_\kappa + r\beta_\kappa']^2 + 2\beta_\kappa^4) - \frac{(\beta_\kappa^2 - 1)^2 + 2r^2 \beta_\kappa'^2 + 2C_M^2 r^4 \beta_\kappa^4}{r^4} \tag{5.5}$$

where the r-dependence of $\beta_\kappa$ is suppressed for simplicity. The CP-odd invariant is given by the expression,

$$(E_i^a B_i^a)^{(1)}(r) = C_E \frac{\beta_\kappa^2(\beta_\kappa - r\beta_\kappa') - (\beta_\kappa + r\beta_\kappa')}{r^2} \tag{5.6}$$

The scalar density, (5.5), for the type-1 solutions has a similar shape to that of the type-0 solutions with a lower maximum due to the larger value of for the magnetic terms. The very important supplement provided by these solutions involves the CP-odd density given by (5.6). A sketch of this density is shown in Fig. (3). The CP-odd density is proportional to $C_E$ and does not depend on the constant $C_M$ because in the field-strength formulation this constant only appears in the expression for $B_A$ and the static expression for $E_A$ vanishes. The topological structure provided by a domain wall consisting of a CP-odd volume of space provides stability to these solutions. Recall that the exterior volume in these solutions does not describe a classical vacuum since it contains a radial magnetic field-strength.

The field strengths in the transition region for type-2 solutions are given by:

$$E_L(r) = C_E[\beta_\kappa(r) + r\beta_\kappa'(r)] \quad E_S(r) = C_E\alpha_\kappa(r)\beta_\kappa(r) \quad E_A(r) = 0.$$

$$B_L(r) = \frac{\alpha_\kappa^2(r)-1}{r^2} \qquad B_S(r) = \frac{-\alpha_\kappa'(r)}{r} \qquad B_A(r) = \frac{\alpha_\kappa(r)}{r}[C_M\beta_\kappa(r) - \pi\beta_\kappa'(r)].$$

(5.7)

These expressions lead to the scalar invariant density,

$$(E_i^a E_i^a - B_i^a B_i^a)^{(2)}(r) = C_E^2([\beta_\kappa + r\beta_\kappa']^2 + 2\beta_\kappa^2\alpha_\kappa^2) + \frac{(\alpha_\kappa^2-1)^2 + 8r^2\beta_\kappa'^2 + 2r^2\alpha_\kappa^2[C_M\beta_\kappa - \pi\beta_\kappa']^2}{r^4}$$

(5.8)

where we have simplified the expression by using $\alpha_\kappa'(r) = 2\beta_\kappa'(r)$ and omitted the arguments of the functions. The CP –odd invariant density is then given by

$$(E_i^a B_i^a)^{(2)}(r) = C_E \frac{[\beta_\kappa + r\beta_\kappa'](\alpha_\kappa^2-1) - 4r\beta_\kappa\alpha_\kappa\beta_\kappa'}{r^2}$$

(5.8)

A simple comparison of the three different scalar densities in the transition region is sketched in Fig. X. The two nonzero pseudoscalar densities are compared in Fig. X+1.

It also important to consider that the interpolations considered here will serve to generate dynamically-induced color charges in the transition region in addition to those currents that interpolate from the interior region. This transition region, therefore can be described as a color-riparian zone in addition to its other dynamical properties. The currents consistent with the static Yang-Mills Maxwell equations (4.2) for the type-0 interpolation are given by

$$J_0^{(0)}(r) = -C_E r^2 \left(4\beta_\kappa' + r\beta_\kappa''\right)$$

$$J_1^{(0)}(r) = 2C_M r\beta_\kappa$$

$$j_S^{(0)}(r) = \frac{C_M}{r}(\beta_\kappa + r\beta_\kappa')$$

$$j_A^{(0)}(r) = -(C_E^2 - C_M^2)r\beta_\kappa^2$$

(5.9)

Because $a(r) = 1$ in both the interior and exterior volumes as well as in the transition zone for the type-0 interpolations, small fluctuations in the field equations can allow the solutions to spread beyond $r = R_0 + \Delta$ for these solutions. The situation is much different for the type-1 solutions.

The color currents in the transition region consistent with the static Yang-Mills Maxwell equations for the type-1 interpolations can be written

$$J_0^{(1)}(r) = -C_E[2r\beta_\kappa(1-\beta_\kappa^2) + 4r^2\beta_\kappa' + r^3\beta_\kappa'']$$
$$J_1^{(1)}(r) = 2C_M r\beta_\kappa^3$$
$$j_S^{(1)}(r) = \frac{C_M}{r}(\beta_\kappa^2 + 3r\beta_\kappa\beta_\kappa') \tag{5.10}$$
$$j_A^{(1)}(r) = \frac{1}{r}[\beta_\kappa'' - (C_E^2 - C_M^2)r^2\beta_\kappa^3 - \frac{\beta_\kappa(\beta_\kappa^2-1)}{r^2}]$$

With $a(r) = 0$ in the exterior region, the transverse field-strength densities that carry color are explicitly repelled from the region beyond $r = R_0 + \Delta$. The pillow of color charges in the transition region also cannot intrude there.

The color currents in the transition region for the type-2 interpolations given above are specified by

$$J_0^{(2)}(r) = -C_E[2r\beta_\kappa(1-\alpha_\kappa^2) + 4r^2\beta_\kappa' + r^3\beta_\kappa'']$$
$$J_1^{(2)}(r) = 2\alpha_\kappa^2(C_M r\beta_\kappa - \pi\beta_\kappa')$$
$$j_S^{(2)}(r) = \frac{4\beta_\kappa'}{r}[C_M r\beta_\kappa - \pi\beta_\kappa'] + \frac{\alpha_\kappa}{r}[C_M(\beta_\kappa + r\beta_\kappa') - \pi\beta_\kappa''] \tag{5.11}$$
$$j_A^{(2)}(r) = \frac{1}{r}[\alpha_\kappa'' + \alpha_\kappa(C_M r\beta_\kappa - \pi\beta_\kappa')^2 - r^2\alpha_\kappa C_E^2\beta_\kappa^2 - \frac{\alpha_\kappa(\alpha_\kappa^2-1)}{r^2}]$$

Unlike the type-1 interpolations, there is no additional barrier to repel color from the $a(r) = -1$ vacuum in the interior region. What distinguishes these solutions is that the covariant derivatives change sign when $a(r)$ changes sign. In the type-2 interpolations this occurs at the surface defined by $r = R_0$. There is much work that remains to be done in testing the properties of both type-1 and type-2 solutions. Preliminary steps in this process may involve consideration of the time-dependent solutions of (2.19) that describe small fluctuations around the time-independent solutions considered in this section. In addition, the time-dependent solutions to the Weyl-Dirac equations (3.18) in the background of the gauge fields specified here.

Extending the topological content of type-1 and type-2 solutions considered here to the gauge group SU(3) can be attempted with the aid of a 2-dimensional charge space $H = (Q_3, Q_8)$ and root vectors, $\omega_a$, so that the six non-diagonal generators of SU(3), $E_{\pm a}$, can be expressed by the commutators

$$[H, E_{\pm a}] = \pm\omega_a E_{\pm a}, \quad (a = 1, 2, 3) \tag{5.12}$$

allows for the combination of spherical symmetry and the maximal Abelian gauge. This formalism was discussed by Ripka [2] in the context of dual superconductor models. Obviously, the algebraic constraints in the SU(3) formalism will be challenging.

Section 6. Constructive Quantum Field Theory.

Classical field theory can provide direct insight into quantum field theory in many ways. In particular, it is very interesting when a classical field theory has solutions that are "particle-like" at the classical level. As pursued in detail in Erick Weinberg's book [45], the analysis of solutions to classical field equations that are "localized in space and do not dissipate over time" leads to the important question of whether such solutions have counterparts in the quantum theory that are quite distinct from the wave solutions describing the fundamental constituents of perturbation theory. The type-1 and type-2 solutions to the Yang-Mills Maxwell equations for and SU(2) gauge theory presented above in the limit of spherical symmetry, clearly achieve this threshold of interest for the description of quantum confinement. The spherical domain wall of topological charge separating an interior volume with classical SU(2) charges and an exterior volume with vacuum properties found in the type-2 solutions can be placed in direct analogy to the soliton or "kink" solutions in the 1+1 dimensional Abelian Higgs model. The review article of Frohlich [46] outlines the formal process of constructive field theory pioneered by Bridges, Frohlich and Seiler [47] for this model that leads to a mass gap in the corresponding continuum field theory. For type-1 solutions the topological charge in the transition volume can be considered the surface pillow separating the interior volume of SU(2) charge from the exterior volume with a radial magnetic field. The exterior region with $a(r,t) = 0$ explicitly repels the color-charged transverse fields. The confinement of the color-neutral radial electric field, $E_L(r,t)$, is, however, independent of $a(r,t)$ and, in this classical formalism, must be put in by hand. Note that with $a(r,t) = 0$ the orientation of the gauge group is not defined since, by eq. (2.5), the covariant derivative of $\hat{r}_a$ and all of its commutators must vanish. This means that the radial magnetic field is dynamically equivalent to an exterior magnetic condensate.

Since the layer of topological charge is strongly connected to the pion tornado of chiral dynamics, the study of the SU(2) Weyl-Dirac equation with a background gauge field obeying type-1 boundary conditions can provide insight into a possible chiral condensate in the exterior volume.

For type-2 solutions presented here, the topological charge density $E_i^a B_i^a$ contains an additional zero associated with the change in sign of the covariant derivatives in (2.5) as $a(r,t)$ passes through zero. For the specific parameters considered here there are also two radial bands in the transition volume where the scalar density $E_i^a E_i^a - B_i^a B_i^a$ is negative indicating a local dominance of color magnetic fields. Unlike the type-1 solutions, the confinement of the transverse color fields is not automatic and it is important to study further the dynamical implications of small fluctuations in order to understand the stability of the classical picture in the exterior region. The comparison of small fluctuations for type-1 and type-2 solutions may prove informative in the discussion of in-hadron condensates [12] and vacuum condensates [13].

The application of spherical symmetry provides some welcome simplicity for the study of the surface color gradients required by confinement. Relaxing this requirement my introduce more dynamical complexity but it is possible that these complications can also be managed.

### FIGURE CAPTIONS

Fig. 1  To study the classical SU(2) dynamics of a confined system with spherical symmetry, it is convenient to define and interior volume $r \leq R_0 - \Delta$; an intermediate, transition volume $r \subset (R_0 - \Delta, R_0 + \Delta)$; and an exterior volume $r \geq R_0 + \Delta$.

Fig. 2  Sketches showing the scalar color density $E_i^a E_i^a - B_i^a B_i^a$ in the transition region for type-0, type-1 and type-2 solutions to the SU(2) Yang-Mills Maxwell equations with spherical symmetry.

Fig. 3  Sketches showing the CP-odd SU(2) color density $E_i^a B_i^a$ for type-1 and type-2 solutions to the SU(2) Yang-Mills Maxwell equations with spherical symmetry.

# SU(2) Color Confinement

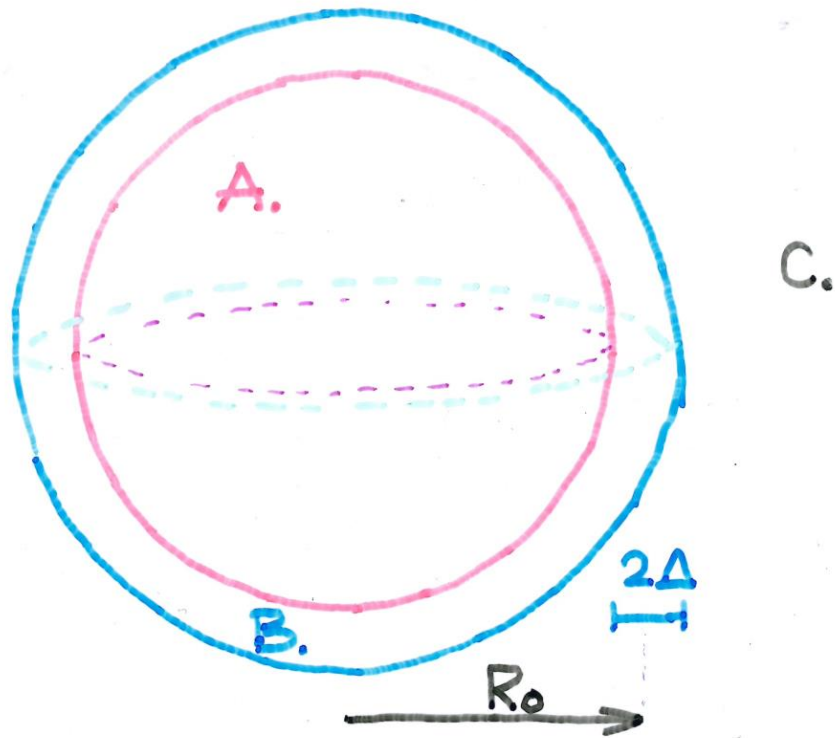

A. Interior Volume - SU(2) charges
B. Transition Volume - color gradients
C. Exterior Volume - no SU(2) charge

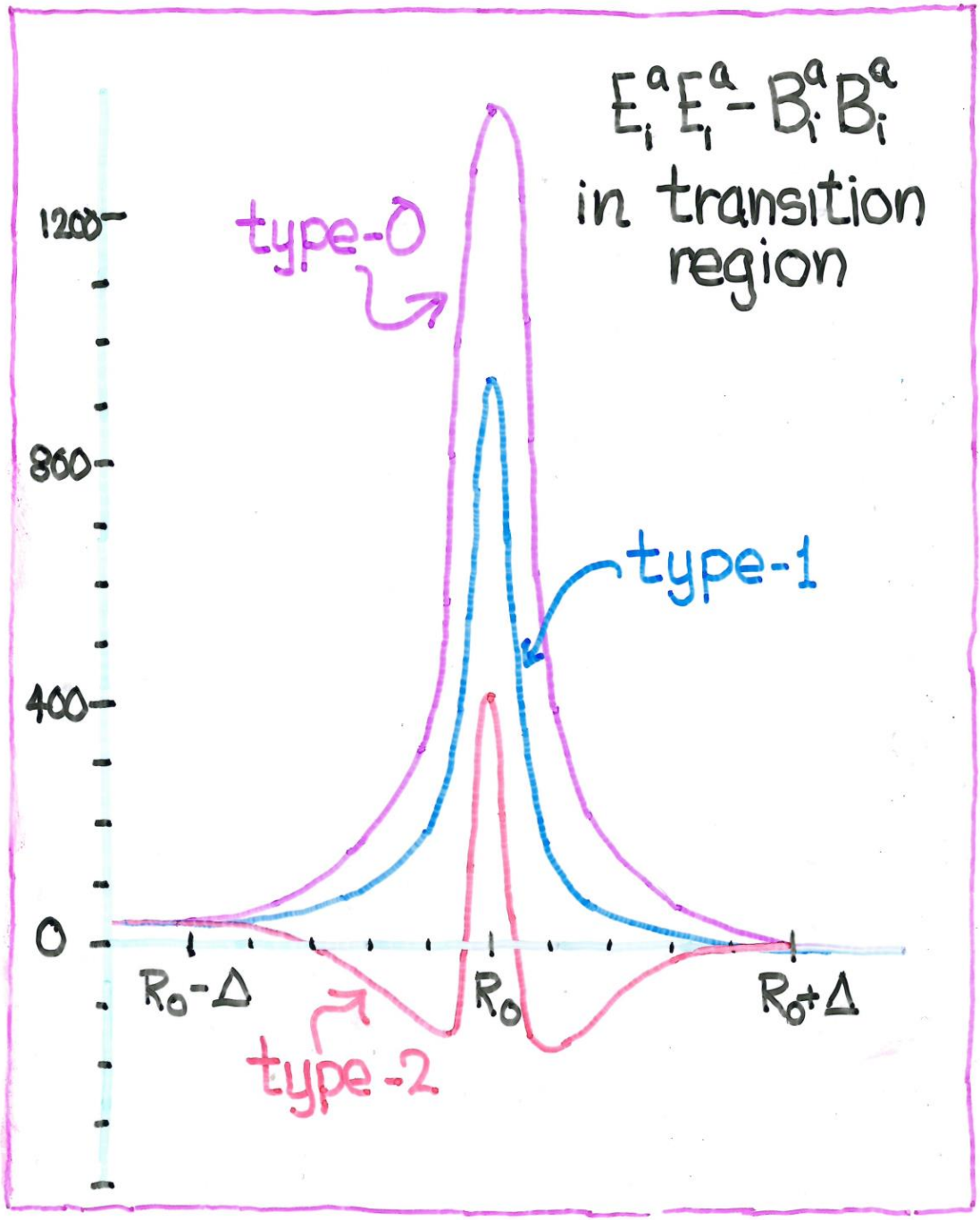

fig. 2

fig. 3

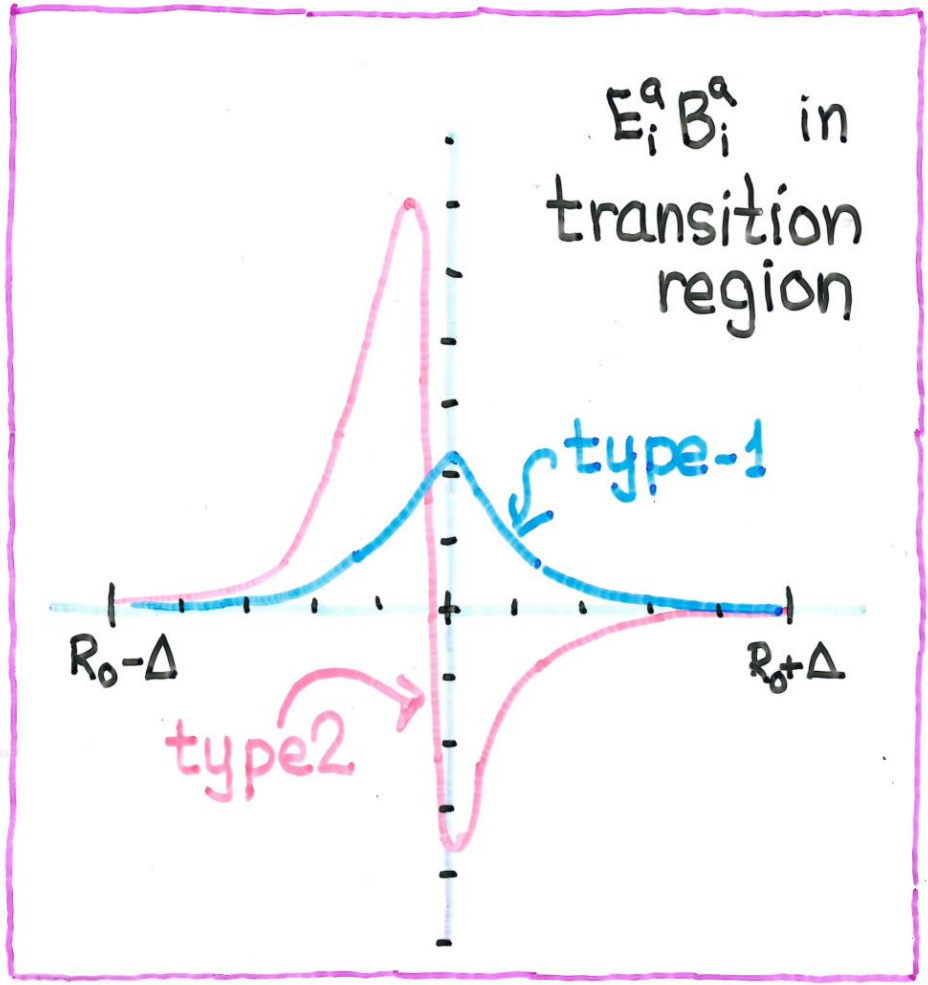